\newlength{\extralineskip}
  \documentstyle[12pt]{article}
\makeatletter
\newdimen\normalarrayskip              
\newdimen\minarrayskip                 
\normalarrayskip\baselineskip
\minarrayskip\jot
\newif\ifold             \oldtrue            

\newcommand{\beq}{\begin{equation}}
\newcommand{\eeq}[1]{\label{#1}\end{equation}}
\newcommand{\beqn}{\begin{eqnarray}}
\newcommand{\eeqn}[1]{\label{#1}\end{eqnarray}}

\newcommand{\ds}{\!\!\not\!\partial}
\newcommand{\as}{\not\!\! a}
\newcommand{\Bs}{\not\!\! B}
\newcommand{\Cs}{\not\!\! C}
\newcommand{\bs}{\not\!\! b}

\newcommand{\D}{{\cal{D}}}

\newcommand{\Z}{{\cal Z}}

\begin{document}
\begin{titlepage}
\setcounter{footnote}0
\begin{center}
\hfill La Plata-Th 96/05\\
\hfill CCNY-HEP-96/5\\
\hfill hep-th/9603138\\
\vspace{0.3in}
{\LARGE\bf Duality in deformed coset fermionic models}
\\[.4in]
{\Large Daniel C. Cabra\footnote{CONICET, Argentina.
E-mail address: cabra@venus.fisica.unlp.edu.ar}}

\bigskip {\it Departamento de F\'\i sica,  \\
Universidad Nacional de La Plata,\\
C.C. 67 (1900) La Plata, Argentina.}\\
and\\
\bigskip
{\Large Enrique F. Moreno\footnote{E-mail address:
moreno@scisun.sci.ccny.cuny.edu}}\\
\bigskip {\it Department of Physics, \\
City College of New York, \\
New York, NY 10031, U.S.A.}
\bigskip\\
\end{center}
\bigskip

\centerline{\bf ABSTRACT}
\begin{quotation}
We study the $SU(2)_k/U(1)$-parafermion model perturbed by its first
thermal operator.  By formulating the theory in terms of a (perturbed)
fermionic coset model we show that the model is equivalent to
interacting WZW fields modulo free fields.  In this scheme, the order
and disorder operators of the $Z_k$ parafermion theory are constructed
as gauge invariant composites. We find that the theory presents a duality
symmetry that interchanges the roles of the spin and dual spin operators.
For two particular values of the coupling constant we find that the theory
recovers conformal invariance and the gauge symmetry is enlarged.
We also find a novel self-dual point.
\end{quotation}
\end{titlepage}
\clearpage

\newpage


\vspace{1cm}

{\it i) Introduction}

\vspace{1cm}

\setcounter{footnote}{0}

In the present article we reformulate the $SU(2)_k/U(1)$ ($Z_k$)
parafermion model \cite{FK,FZ} perturbed by its first thermal operator
\cite{F,FZ2},
in terms of a constrained fermionic model. This model has recently
attracted much attention \cite{Bakas,Park,PS} in connection with the
complex sine-Gordon theory, and the study of classical solutions has
been extensively developed.

Although this model has been widely studied, almost all the work done
about it
dealt with its classical aspects. It is the purpose of the present note
to give a suitable scenario to study its quantum aspects. We show that
the formulation of the model using constrained fermionic models is
particularly useful for this purpose. In particular, the connection
between the perturbed $Z_k$ parafermion model and a Gross-Neveu-like
model is made apparent. We also construct the $Z_k$ primaries in terms
of composite fermions, thus giving an explicit realization of the
Fateev-Zamolodchikov parafermion algebra \cite{FZ}.

The perturbed system presents a remarkable duality transformation: the
theories with coupling constants $\beta$ and $-\beta/(1-2(1+2k)\beta)$
are equivalent but the roles of the spin fields $\sigma$ and the dual
spin fields $\mu$ are interchanged. This transformation is self-dual at
the point $\beta=1/(2k+1)$.

We also show the existence of two non-trivial fixed points (mutually related
by the duality symmetry), where
conformal invariance is explicit and the non-abelian symmetry, originally
$SU(k)$, is enlarged to $SU(k) \times SU(k)$.

\vspace{1cm}

{\it ii) $SU(2)_k$ Wess-Zumino-Witten (WZW) theory as a fermionic coset}

\vspace{1cm}

To start with, let us recall the fermionic coset formulation of the
$SU(2)_k$ WZW theory \cite{NS,BRS,nos}. The action is given by
\beq
S_0=-\frac{1}{\sqrt 2\pi} \int d^2 x
\psi^{\dagger i \alpha}\left( (i\ds +i\as
)\delta_{ij}\delta_{\alpha\beta}+
i\delta_{ij}\Bs_{\alpha\beta}
\right)
\psi^{j\beta} ,
\eeq{1}
where the fermions $\psi^{i\alpha}$ ($i=1,2$, $\alpha =1,...,k$) are in
the fundamental representation of $U(2k)$ and the $U(1)$ gauge field
$a_{\mu}$ and the $SU(k)$ gauge field $B_{\mu}$
act as Lagrange multipliers implementing the
constraints
\beq
j_{\mu}|phys>=0 ,~~~~~~~~J_{\mu}^a|phys>=0
\eeq{2}
for $j_{\mu}$ the $U(1)$ and $J_{\mu}^a$ the $SU(k)$ currents
respectively.
This corresponds to the identification:
\beq
SU(2)_k \equiv \frac{U(2k)}{SU(k)_2 \times U(1)} ,
\eeq{3}
which is understood as an equivalence between the correlation functions
of corresponding fields in the two theories.

The fundamental field $g$ and its adjoint $g^{\dagger}$ of the
bosonic $SU(2)_k$
WZW theory are represented in terms of fermions by the bosonization
formulae\footnote{
Our conventions are $\psi=\left(\begin{array}{l}
\psi_1 \\
\psi_2
\end{array}\right)$ and $\gamma_i$ are the Pauli matrices.}
 \cite{NS}

\beqn
g^{ij}&=&\psi_2^i\psi_2^{j\dagger}
\nonumber \\
g^{ij\dagger}&=&\psi_1^i\psi_1^{j\dagger} ,
\eeqn{4}
where the $SU(k)$ indices are summed out.

Higher spin integrable representations are constructed as symmetrized
products of these fundamental fields

\beqn
g^{(j)~i_1,...,i_{2j}}_{j_1,...,j_{2j}}&=&{\cal S} \left(:g^{i_1}_{j_1}
...g^{i_{2j}}_{j_{2j}}:\right)
\nonumber \\
&=&{\cal S} \left(:\psi_2^{i_1}\psi_2^{\dagger j_1}
...\psi_2^{i_{2j}}\psi_2^{\dagger j_{2j}}:\right),
\eeqn{4'}
where $j=0,1/2,1,...,k/2$. This restriction in the spin of the
representation has its origin in the selection rules imposed by
the Kac-Moody symmetry \cite{GW,FGK}.

In the fermion representation, the presence of a second index $\alpha$
in
the fermion fields $\psi^{i\alpha}$, running from $1$ to $k$, allows
for the construction of symmetrized products of at most $k$ bilinears.
In this way we obtain only the allowed integrable representations of
spin $j=0,1/2,1,...,k/2$
and higher spin representations are forbidden by the Pauli principle.

Before going to the perturbed case, we will first construct the
partition function of the $Z_k$ parafermion theory
as the fermionic coset $SU(2)_k/U(1)$ and construct
the primary fields in terms of the constrained fermions.
The construction of these primary fields
is one of our new results. As far as we know, an
explicit
representation of the primary fields in $Z_k$ parafermion models
starting from first principles has not been presented before,
although some work related to classical aspects has been done
\cite{BCR}.

\vspace{1cm}

{\it iii) $Z_k$ formulated as the coset $SU(2)_k/U(1)$}

\vspace{1cm}

To carry out this program we have to further mod out a $U(1)$ subgroup
of the unconstrained $SU(2)$ group in the $SU(2)_k$ theory (\ref{1}).
We will do this by freezing the component of the
$SU(2)$ current in the direction of $t^3$,

\beq
j_{\mu}^3=:\psi^{i\alpha\dagger}\gamma_{\mu}t^3_{ij}\psi^{j\alpha}:.
\eeq{5}
To implement this constraint in the path-integral formulation we have to
add to the action a term of the form
\beq
-\frac{i}{\sqrt{2} \pi} \int d^2 x
\psi^{i\alpha\dagger}\bs^3 t^3_{ij}\psi^{j\alpha} ,
\eeq{6}
and integrate over the gauge field $b_{\mu}^3$, thus obtaining
\beq
{\cal{Z}}_{SU(2)_k/U(1)}=\int Da_{\mu} DB_{\mu}Db_{\mu}^3
D\psi^{\dagger}D\psi e^{-{\cal S}} ,
\eeq{6'}
where
\beq {\cal S} = \frac{1}{\sqrt 2 \pi} \int d^2 x
\psi^{\dagger i \alpha}\left( (i\ds +i\as
)\delta_{ij}\delta_{\alpha\beta}+
i\delta_{ij}\Bs_{\alpha\beta}+i\bs^3 t^3_{ij}\delta_{\alpha \beta}
\right)
\psi^{j\beta} .
\eeq{6a}
The Virasoro central charge of this model can be evaluated using
standard methods (see the appendix) and is given by
\beq
c=\frac{2(k-1)}{k+2} ,
\eeq{cc}
which corresponds to the central charge of the $Z_k$-parafermion theory.

The parafermion models and their integrable perturbations have been
widely studied\\ \cite{FZ,F,FZ2} using Conformal Field Theory
techniques.
It was shown in particular that the primary fields of the
$Z_k$-parafermion theory, $\phi^{(j)}_{m,\bar m}$,
are closely related to the primaries of the
$SU(2)_k$ WZW theory, $\Phi^{(j)}_{m,\bar m}$ \cite{FZ}.
Indeed, they are connected by the relation
\beq
\Phi^{(j)}_{m,\bar m}(z,\bar z)=
\phi^{(2j)}_{2m,2\bar m}(z,\bar z) :\exp \left(i(\frac{m}{\sqrt{k}}
\varphi(z)+
\frac{\bar m}{\sqrt{k}} \bar{\varphi} (\bar z))\right): ,
\eeq{fz}
where $\varphi(z)$ and $\bar{\varphi}(\bar z)$ are the holomorphic and
antiholomorphic components of an auxiliary
massless free boson field.

However, although this relation has proven to be useful, it should be
desirable to have an explicit representation of the primaries in
the $Z_k$-parafermion models.

Let us first recall which are the more relevant primaries in
$Z_k$ parafermion models.

Among the primaries $\phi^{(j)}_{m,\bar m}$, there are the ``spin''
fields $\sigma_p=\phi^{(p)}_{p,\bar p}$ and the ``dual spin'' fields
$\mu_p=\phi^{(p)}_{p,-\bar p}$, with conformal dimensions
\beq
h_p=\bar{h}_p=\frac{p(k-p)}{2k(k+2)}
\eeq{dimen}
where the duality (Krammers-Wannier) symmetry corresponds to the
interchange
\beq
\sigma \leftrightarrow \mu.
\eeq{kw}

Other relevant primaries are the so-called parafermion currents,
$\Psi^{PF}_p=\phi^{(0)}_{2p,0}$ and
${\bar{\Psi}}^{PF}_p=\phi^{(0)}_{0,2p}$, $p=1,2...,k-1$, with
conformal dimensions $(h, \bar h)$ given by
\beq
(\frac{p(k-p)}{k}, 0) \ \ \ \ {\rm and} \ \ \ \ (0, \frac{p(k-p)}{k})
\eeq{dimpf}
respectively and the ``thermal'' operators,
$\epsilon_j=\phi^{(2j)}_{0,0}$, $j=1,2,...\le k/2$, with dimensions
$h_j=\bar h_j= j(j+1)/(k+2)$. These thermal operators are the only ones
explicitly represented by eq.(\ref{fz}) in terms of the WZW $SU(2)_k$
primaries.

We will now show how these primary fields can be constructed as gauge
invariant
fermion composites in the theory described by the partition function
(\ref{6'}).

In order to identify the physical fields in this theory, we have to seek
for gauge invariant operators, which will be constructed using the gauge
invariant fermions \cite{CR1}.

The fields in (\ref{4}) and (\ref{4'}) are already gauge invariant under
gauge transformations associated with the gauge fields $a_{\mu}$ and
$B_{\mu}$, but they vary under gauge transformations associated with the
new gauge field, $b_{\mu}^3$, introduced in eq.(\ref{6}).  In order to
ensure invariance also under these transformations, we will define
the gauge invariant fermions as

\beq
\hat{\psi}=e^{-i\int_x^{\infty} dz^{\mu}b_{\mu}^3}\psi .
\eeq{gif}

Using these fields we can construct the gauge invariant version of the
$g$-field and its adjoint in eq.(\ref{4}) \cite{CR2}
\beqn
{\hat g}_{ij}&=&{\hat{\psi}}_2^i{\hat{\psi}}_2^{j\dagger},
\nonumber \\
{\hat g}_{ij}^{\dagger}&=&{\hat{\psi}}_1^i{\hat{\psi}}_1^{j\dagger}
\eeqn{bb}
and using as a guideline the identification made in ref.\cite{FZ},
i.e. eq.(\ref{fz}), we will study the fields
\beqn
\sigma_1&\equiv & {\hat g }_{1,1} + {\it h.c.}\ ,
\nonumber \\
\mu_1&\equiv & {\hat g }^{\dagger}_{2,1} + {\it h.c.} 
\eeqn{sm}
These fields have dimensions (see the appendix)
\beq
h=\bar{h}=\frac{k-1}{2k(k+2)}
\eeq{dim}
and satisfy the dual algebra \cite{CR1}
\beq
\sigma_1(x_1) \mu_1(x_2) =e^{\frac{i2\pi}{k}\Theta(x_1-x_2)}
\mu_1(x_2)\sigma_1(x_1).
\eeq{od}
Eqs. (\ref{dim}) and (\ref{od}) lead one to identify the fields $\sigma$
and $\mu$ defined in (\ref{sm}) with the order and disorder operators in
the $Z_k$ parafermion theory.

The other spin fields, ($p>1$),
are obtained by using eq.(\ref{4'}) but with the
gauge
invariant fermions, and following the same lines as above, it can be
shown that the fields
\beqn
\sigma_p&\equiv &
:\underbrace{{\hat g}_{1,1}{\hat g}_{1,1}\ldots {\hat g}_{1,1}}_{p-times}
: +\ {\it h.c.}\ ,
\nonumber \\
\mu_p&\equiv &
:\underbrace{{\hat g}^{\dagger}_{2,1}{\hat g}^{\dagger}_{2,1}\ldots
{\hat g}^{\dagger}_{2,1}}_{p-times}
:  +\ {\it h.c.}\ ,
\eeqn{smg}
have dimensions given by eq.(\ref{dimen}) and satisfy the algebra
\beq
\sigma_p(x_1) \mu_{p'}(x_2) =e^{\frac{i2pp'\pi}{k}\Theta(x_1-x_2)}
\mu_{p'}(x_2)\sigma_p(x_1) ,
\eeq{odg}
as required.

The Krammers-Wannier symmetry corresponds in this context to the
interchange of the fermion components:

\beq
\psi^{1 \alpha}_2 \leftrightarrow \psi^{1 \alpha \dagger}_1\ ,\ \ \ \ \ \ \ \ \
\psi^{1 \alpha \dagger}_2\leftrightarrow \psi^{2 \alpha}_1.
\eeq{6''}
Using eq.(\ref{bb}) we can write the duality transformation as
\beq
{\hat g}_{ij} \rightarrow \left({\hat g} \gamma_1\right)^{\dagger}_{ij},
\eeq{6'''}
that resembles the one proposed in ref.\cite{PS}
but involves the gauge invariant version of the $SU(2)$ field and
is valid for any $k$.

Besides the bilinears defined in eq.(\ref{bb}), we can construct another
class of bilinears in terms of the gauge invariant fermions,
which have no local counterpart in
the bosonic formulation. Let us consider the fields
\beq
{\hat \psi}_1^2 {\hat \psi}_2^{1\dagger}\ \ ,\ \ \ \ \ \ \ \ \
{\hat \psi}_2^2 {\hat \psi}_1^{1\dagger}.
\eeq{pf}
These fields have conformal dimensions $(h, \bar h)$ (see the appendix)
\beqn
(\frac{k-1}{k}, 0)\ \ \ \ {\rm and} \ \ \ \ (0, \frac{k-1}{k}),
\eeqn{dimpf'}
respectively and then can be identified with the basic parafermion
currents,
\beq
{\Psi}_1^{PF}={\hat{\psi}}_1^2 {\hat{\psi}}_2^{1\dagger}\ +\ {\it h.c.}
\ \ \ \ \ \ {\rm and} \ \ \ \ \
{\bar{\Psi}}_1^{PF}={\hat{\psi}}_2^2 {\hat{\psi}}_1^{1\dagger}\ +
\ {\it h.c.}
\eeq{pfc}

The other parafermion currents are constructed as
\beqn
\Psi_p^{PF}&\equiv &
:\underbrace{{\hat{\psi}}_1^2
{\hat{\psi}}_2^{1\dagger}{\hat{\psi}}_1^2
{\hat{\psi}}_2^{1\dagger}...{\hat{\psi}}_1^2
{\hat{\psi}}_2^{1\dagger}}_{p-times}
: +\ {\it h.c.}\ ,
\nonumber \\
{\bar\Psi}_p^{PF}&\equiv &
:\underbrace{{\hat{\psi}}_2^2
{\hat{\psi}}_1^{1\dagger}{\hat{\psi}}_2^2
{\hat{\psi}}_1^{1\dagger}...{\hat{\psi}}_2^2
{\hat{\psi}}_1^{1\dagger}}_{p-times}
: +\ {\it h.c.}\ ,
\eeqn{pfg}
with dimensions given respectively by
\beq
(\frac{p(k-p)}{k}, 0)\ \ \ \ {\rm and} \ \ \ \ (0, \frac{p(k-p)}{k}).
\eeq{dimpfg}

Finally, let us stress that the order and disorder operators and the
parafermionic currents, as defined by eqs.(\ref{smg}) and (\ref{pfg})
respectively,
satisfy the conformal operator product expansions:

\beq
\begin{array}{l}
\sigma_p(z,\bar z) \mu_p(0,0) \propto
z^{\frac{p(k-p)(k+1)}{k(k+2)}}{\bar z}^{-\frac{p(k-p)}{k(k+2}}
  \left(\Psi^{PF}_p(0)\ +\ O(z)
\right),
\nonumber\\
\Psi^{PF}_p(z) \Psi^{PF}_q(0) \propto z^{-\frac{2 p q}{k}} \left(
\Psi^{PF}_{p+q}(0)\ +\ O(z) \right),
\end{array}
\eeq{cope}
as expected \cite{FZ}. These operator expansions are easily
calculated in the present formulation by
using the decoupled approach (see the appendix).

\newpage

\vspace{1cm}

{\it iv) Perturbed $Z_k$ parafermion model}

\vspace{1cm}

We are now going to study the model discussed
above perturbed by the primary
operator
corresponding to the first thermal operator,
$\epsilon_1=\phi_{00}^{(2)}$ \cite{FZ}.
In terms of WZW fundamental fields $\epsilon_1$ can be written as (see
eq.(\ref{fz})):

\beq
\epsilon_1=Tr \left(:gt^3g^{\dagger}\bar{t}^3:\right) \ ,
\eeq{7}
and has conformal dimensions $h={\bar h}=2/(2+k)$.

Using the bosonization dictionary (\ref{4}), the interaction term can be
written as:
\beq
S_{int}=-\frac{\beta}{4\pi} \int d^2 x Tr \left(
K_{z}^3 K_{\bar z}^3 \right) ,
\eeq{8}
where ${K_{\mu}^3}^{\alpha \beta}=\psi^{i \alpha \dagger}\gamma_{\mu}t^3
\psi^{i \beta}$.
Note that in the present case the gauge invariant fields
introduced in eq.(\ref{bb})
are not needed since the expression for $\epsilon_1$ is already gauge
invariant in terms of the $SU(2)_k$ fields.

At this stage we want to point out the connection between the $Z_k$
parafermion model and a Gross-Neveu-like model which is apparent in the
fermionic coset formulation. Indeed, the whole action
$S_0+S_{int}$, (eqs.(\ref{6a}) and (\ref{8})), which represents the
perturbed parafermion model, corresponds to gauged fermions with an
$SU(2)$ Gross-Neveu-like interaction term. The connection between the
perturbed parafermion model and a Gross-Neveu-like model has been
suggested \cite{Bakas}.

We now introduce an auxiliary field $C_{\mu}^3$ through the identity
\beq
\exp \left( \frac{\beta}{4\pi} \int d^2 x Tr (K_{\mu}^3 K_{\mu}^3)
\right)=
\int DC_{\mu}^3 \exp \left(-\frac{1}{\pi \beta}\int d^2 x Tr(C_{\mu}^3)^2
+ \frac{1}{\pi}\int d^2x Tr K_{\mu}^3 C_{\mu}^3 \right)
\eeq{9}
so the action can be written as:
\beqn
S&=&-\frac{1}{\sqrt 2\pi} \int d^2 x
\psi^{\dagger i \alpha}\left( (i\ds +i\as
)\delta_{ij}\delta_{\alpha\beta}+
i\delta_{ij}\Bs_{\alpha\beta}+i\bs^3 t^3_{ij}
\delta_{\alpha\beta}-(\Cs^3)^{\beta\alpha}t^3_{ij}\right)
\psi^{j\beta}\nonumber \\
&+& \frac{1}{\pi \beta}\int d^2 x Tr (C_{\mu}^3)^2.
\eeqn{10}
Though the auxiliary field $C_{\mu}$ belongs to the Lie algebra of
$U(k)$, its $U(1)$ component decouple from the rest.  In the Dirac
operator it can be absorbed by the field $b^3_{\mu}$ and in the
interaction it just decouples from the fermion fields, leading to a
Gaussian integral which indeed does not modify the conformal algebra.
Therefore
we will simply discard the $U(1)$ piece of $C^3_{\mu}$.

The action (33) is invariant under independent $U(1)$ gauge
transformations
for the fields $a_{\mu}$ and $b_{\mu}$ and also has the local $SU(k)$
symmetry
\beq
\begin{array}{ll}
B_{\mu} \to g^{-1}B_{\mu}g - g^{-1}\partial_{\mu} g\ , \ \ \
&C_{\mu} \to g^{-1} C_{\mu} g,\nonumber \\
\psi^{i} \to g \psi^{i},
&{\psi}^{\dagger i} \to {\psi}^{\dagger i} g^{-1}.
\end{array}
\eeq{10.5}
So the combination $B_{\mu} \pm C_{\mu}$ transforms as a gauge field.
The chiral version of these symmetries are anomalous.
We will see later that for special values of the coupling constant
this symmetry is actually  enlarged and the theory becomes invariant
under a  local $SU(k) \times SU(k)$ symmetry.

The key observation now is that the fields in eq.(\ref{10}) can be almost
completely decoupled through transformations:

\beq
\begin{array}{l}
{\begin{array}{ll}
a_{\mu} = \partial_{\mu} \eta_a - \epsilon_{\mu \nu} \partial_{\nu}
\phi_a\ , &
b_{\mu} = \partial_{\mu} \eta_b - \epsilon_{\mu \nu} \partial_{\nu}
\phi_b  \nonumber\\
\delta_{i j} B_z + t^3_{i j} C_z =\pmatrix{U_+^{-1} \partial_z U_+ & 0\cr
             0 & U_-^{-1} \partial_z U_-\cr}_{i j}, &
\delta_{i j} B_{\bar z} + t^3_{i j} C_{\bar z} =\pmatrix{V_+^{-1}
\partial_{\bar z} V_+ & 0\cr    0 & V_-^{-1}
\partial_{\bar z} V_-\cr}_{i j}
\end{array}}\nonumber\\
\psi_1=\pmatrix{e^{-(\phi_a + \phi_b)-i(\eta_a +\eta_b)} U_+^{-1} & 0\cr
   0 & e^{-(\phi_a - \phi_b) -i (\eta_a - \eta_b)} U_-^{-1}\cr}\chi_1,
\nonumber\\
\psi_2=\pmatrix{e^{(\phi_a + \phi_b)-i (\eta_a +\eta_b)} V_+^{-1} & 0\cr
       0 & e^{(\phi_a - \phi_b) -i (\eta_a - \eta_b)} V_-^{-1}\cr}\chi_2
\end{array}
\eeq{11}
where $\phi_{a,b},\ \eta_{a,b}$ are scalar fields and
$U_{\pm}$ and $V_{\pm}$ are $SU(k)$ matrix valued fields.

Taking into account the jacobians of the above transformations
and those arising from the gauge-fixing procedure \cite{FiPol}, we can
rewrite the whole partition function as a product of decoupled sectors
(see the appendix)
\beq
{\cal{Z}}={\cal{Z}}_{ff}{\cal Z}_{bos}{\cal{Z}}_{gh}{\cal{Z}}_{int},
\eeq{12}
where ${\cal{Z}}_{ff}$ is the partition function for free
fermions, ${\cal Z}_{bos}$ the partition function for free scalar
bosons,
${\cal{Z}}_{gh}$
the ghost partition function arising in the gauge-fixing procedure
and
\beqn
{\cal{Z}}_{int}&=&\int DU_{\pm} DV_{\pm} \exp\left\{(1+2k)\left(
\Gamma [U_+V_+^{-1}]+
\Gamma [U_-V_-^{-1}]\right)\right.
\nonumber \\
&-& \frac{1}{2 \beta \pi} \int d^2 x
Tr\left[\left.\left(U_+^{-1}\partial_z U_+-U_-^{-1}\partial_z U_-\right)
\left(V_+^{-1}\partial_{\bar z} V_+-
V_-^{-1}\partial_{\bar z} V_-\right)\right]\right\},
\eeqn{13}
where $\Gamma [u]$ is the WZW action \cite{Wi}
\beq
\Gamma[u]=
\frac{1}{16\pi} \int d^2x
Tr \left(\partial_{\mu} u \partial^{\mu} u^{-1} \right) +
\frac{1}{24\pi} \int d^3y \epsilon_{ijk}
Tr \left(u^{-1}\partial_i u u^{-1}\partial_j u u^{-1}\partial_k u
\right) .
\eeq{WZW}

Hence, modulo decoupled conformally invariant sectors, we have an
effective theory of interacting WZW fields.

Using the Polyakov-Wiegmann identity:
\beqn
\Gamma[u v^{-1}] = \Gamma[u] + \Gamma[v^{-1}] -
\frac{1}{2\pi} \int d^2x
Tr \left( u^{-1} \partial_z u v^{-1} \partial_{\bar z} v \right)
\eeqn{13.1}
we can rewrite the interaction action in a more symmetric form:
\beq
S_{eff} = - (1- \frac{1}{2 {\tilde \beta}}) (2k+1)\left( \Gamma[U_+
V_+^{-1}] +
\Gamma[ U_- V_-^{-1}] \right) - \frac{1}{2 {\tilde \beta}} (2k+1)
\left( \Gamma[U_+  V_-^{-1}] + \Gamma[U_- V_+^{-1}] \right),
\eeq{13.2}
where ${\tilde \beta} = (2 k + 1) \beta$.
In this form it is obvious that $S_{eff}$ is invariant under the following
local $SU(k)$ transformation
\beqn
&&U_+ \to U_+ g, \ \ \ \ \ \ \ \ U_- \to U_- g,\\
&&V_+ \to V_+ g, \ \ \ \ \ \ \ \ V_- \to V_- g,
\eeqn{13.25}
which is the representation of the gauge symmetry, eq.(\ref{10.5}), on the
$SU(k)$ valued fields.

This last form of the action, consisting in four WZW actions can give
an illusory appearance of manifest conformal invariance.  However it is
not the case since the arguments of the WZW actions are not all
independent fields
but constrained through the relation

\beqn
(U_+ V_+^{-1})^{-1} U_+ V_-^{-1} (U_- V_-^{-1})^{-1} U_- V_+^{-1}  = 1.
\eeqn{13.3}

The interaction action eq.(\ref{13.2}) has an invariance under
the
duality transformation:
\beqn
U_+ &\to&  U_-\cr
U_- &\to& U_+\cr
{\tilde \beta} &\to& - \frac{\tilde \beta}{1 - 2 {\tilde \beta}}.
\eeqn{13.4}
This transformation, modulo hermitian conjugation, interchanges the
$SU(2)$ components of the fermions in consistency with eq.(\ref{6''}).
Thus it corresponds to the Krammers-Wannier duality $\sigma
\leftrightarrow \mu$.  This symmetry has no effects over the
parafermionic currents. Using a bosonic realization of the deformed
parafermionic theory the authors of
refs.\cite{Bakas,PS} found a {\it classical} invariance under the
transformation $\beta \to -\beta$.
However we have to stress that our result is valid at the
quantum level: is an invariance of the partition function of the
theory.  The duality transformation of ref.\cite{Bakas,PS} agrees
with the leading order term of our eq.(\ref{13.4}) for $\tilde \beta$
small.

It is now straightforward to see the existence of two critical points:
At the value of the coupling constant $1/{\tilde \beta}_1=0$,
the effective partition function can be written as:
\beq
{\cal{Z}}_{int}\left\vert_{\beta_1} \right. =
\int DU_{\pm} DV_{\pm} \exp\{-(2 k +1) \left( \Gamma[U_+
V_+^{-1}] + \Gamma [U_- V_-^{-1}]\right)\}  ,
\eeq{14}
which corresponds to a Conformal Field Theory
whose Virasoro central charge is given by $c=2(1+2k)(k^2-1)/(k+1)$

At the value ${\tilde \beta}_2= \frac{1}{2}$, we have its dual
counterpart:
\beq
{\cal{Z}}_{int}\left\vert_{\beta_2} \right.=
\int DU_{\pm} DV_{\pm} \exp\{-(2 k +1) \left(\Gamma [U_+
V_-^{-1}] + \Gamma [U_- V_+^{-1}]\right)\} ,
\eeq{15}
with the same value for the Virasoro charge.

For these two critical values of the coupling constant the system
acquires an extra symmetry. In fact consider for example the first critical
theory, characterized by the coupling $\beta_1$. Its effective action
is invariant under the transformations
\beqn
&&U_+ \to U_+ g, \ \ \ \ \ \ \ \ \ V_+ \to V_+ g,\\
&&U_- \to U_- h, \ \ \ \ \ \ \ \ \ V_- \to V_- h.
\eeqn{16}
where $g$ and $h$ are independent $SU(k)$ elements.

This doubling of the symmetry has an important consequence on the
parafermionic
primary fields. While the order operator $\sigma$ is invariant under this
symmetry,
the disorder operator $\mu$, being proportional to $U_- V_+^{-1}$ is not.
Thus the
symmetry prevent us for the inclusion of disorder operators in the vacuum
expectation values. The order operator $\sigma$ is still a primary field
in the new critical point but with a vanishing conformal dimension.
The properties of the second fixed point are similar but with the roles of
$\sigma$ and $\mu$ interchanged.

Taking into account the remaining degrees of freedom (fermions, scalars
and ghosts, (see the appendix)),  we can  construct the total conformal
algebra of the critical theories. In both cases the Virasoro central charge
vanishes.

A remarkable curiosity is the fact that the system (\ref{13.2}) has
another self-dual point, besides the one
corresponding to the original parafermionic theory, (i.e. $\beta=0$),
which is reached when the coupling constant takes the value
$\beta=\frac{1}{(1+2 k)}$.  At this novel self-dual point the $SU(k)$ valued
effective action reads
\beq
S_{eff}\vert_{SD} = -(k+ \frac{1}{2})\left( \Gamma[U_+ V_+^{-1}] +
\Gamma[ U_- V_-^{-1}] + \Gamma[U_+  V_-^{-1}] + \Gamma[U_- V_+^{-1}]
\right).
\eeq{17}
At first sight this action does not seem to present conformal
invariance. However, this statement needs a deep analysis as interacting
WZW action can present very non-trivial fixed points
\cite{So}.

\vspace{1cm}

{\it v) Conclusions and discussion}

\vspace{1cm}

We have studied in this paper the $SU(2)_k/U(1)$ parafermionic theory in
terms of a fermionic coset model.  This formulation reveals to be
particularly fruitful for the analysis of the conformal properties of
the model.  In particular the order/disorder operators and the
parafermionic currents can be explicitly constructed and they
take a very simple form in terms of gauge
invariant fermionic fields. The computation of the
conformal OPE's is then straightforward in this scheme.

The inclusion of a deformation in the direction of the first thermal
operator can be implemented by perturbing the coset model with a
Gross-Neveu type interaction
\beq
S_{int}= - \frac{\beta}{4 \pi} \int d^2x\
\psi^{\alpha \dagger}\gamma_{\mu}t^3\psi^{\beta} \
\psi^{\beta \dagger}\gamma_{\mu}t^3\psi^{\alpha}.
\eeq{c1}

After some manipulations within the path integral we have factored out
from the whole partition function a fermionic and two scalars free
partition functions.  The remaining degrees of freedom contain all the
information about the interaction and correspond to a system of coupled
WZW actions.  This last system presents a remarkable duality symmetry:
the theories with couplings constants $\beta$ and $- \frac{\beta}{1 -
2\beta (1 + 2k)}$ are identical.  Under this transformation the roles of
the order an disorder operators are reversed, so it corresponds to a
generalization of the usual Krammers-Wannier duality.
Unlike the duality symmetry of refs. \cite{Bakas,PS}, our result
is valid at the quantum level. The result of refs. \cite{Bakas,PS}
coincides with the tree level approximation of eq. (\ref{13.4}).

We also found two critical values of the coupling constant for
which the theory recovers conformal invariance,
\beq
\beta_1=\frac{1}{2(1+2 k)} \ \ \ \ \ \ {\rm and}\ \ \ \ \
\beta_2=\infty \ \ \ {\rm (its\ dual\ counterpart)}.
\eeq{c2}
For both critical actions the original symmetry is enlarged and the system
is invariant under an extra local $SU(k)$ symmetry.
At both fixed points the theory represents a CFT with total central
charge equal to $0$.

A question that remains open is the evolution of the deformed parafermionic
theory under the renormalization group.
The perturbation from the parafermionic theory is in a relevant
direction (the operator responsible of the deformation has dimension
$\frac{4}{k+2} <2 $).  Thus the perturbation drives away the theory from
the critical point along an infrared renormalization group trajectory.
As the central charge of the parafermionic theory is greater than zero
we can expect, invoking Zamolodchikov's {\sl c}-Theorem, that the model
evolves under the renormalization group to the conformal points
parametrized by $\beta_1$ and $\beta_2$, depending on the sign of the
original perturbation.  The proof of this conjecture is a very
interesting problem though highly non trivial due to its non
perturbative nature.  A perturbative expansion for sufficiently large
$k$ could bring some light to the answer, as one of the fixed points is
of order $1/k$ (but note however that the central charges are finitely
separated for any $k$).  Work in this direction is in progress.

\vspace{.5cm}

{\it Acknowledgements}: D.C. would like to thank G. Rossini for
a careful reading of the manuscript.

\vspace{1cm}

{\it Appendix}
\renewcommand{\theequation}{A.\arabic{equation}}
\setcounter{equation}{0}

\vspace{1cm}
In this appendix we will evaluate the central charge and construct the
primary fields in the $Z_k$-parafermion theory formulated as a fermionic
coset.

The partition function is given by eqs.(\ref{6'}) and (\ref{6a}) and it
can be decoupled through transformations

\beq
\begin{array}{ll}
a_{\mu} = \partial_{\mu} \eta_a - \epsilon_{\mu \nu} \partial_{\nu}
\phi_a &
b_{\mu} = \partial_{\mu} \eta_b - \epsilon_{\mu \nu} \partial_{\nu}
\phi_b  \nonumber\\
B_{z}= 2 i (\partial_{z} U)U^{-1} &
B_{\bar z}= 2 i (\partial_{\bar z} V) V^{-1}  \nonumber\\
\psi_1=e^{-\phi_a - i \eta_a} e^{(-\phi_b - i \eta_b) t_3} U \chi_1 &
\psi_2^{\dagger}=\chi_2^{\dagger}\ U^{-1} e^{\phi_a + i \eta_a}
e^{(\phi_b + i \eta_b) t_3} \nonumber\\
\psi_2=e^{\phi_a - i \eta_a} e^{(\phi_b - i \eta_b) t_3} V
\chi_1 &
\psi_1^{\dagger}=\chi_1^{\dagger}\ V^{-1} e^{-\phi_a + i \eta_a}
e^{(-\phi_b + i \eta_b) t_3},
\end{array}
\eeq{3.d}
where $B_{z}=\frac{1}{2}(B_0 - i B_1)$, $B_{\bar z}=\frac{1}{2}(B_0 +
i B_1)$.

Taking into account the gauge fixing procedure and the Jacobians
associated to (\ref{3.d}) \cite{FiPol} one arrives at
the desired decoupled form for the partition function:
\beq
\Z=\Z_{ff}\Z_{fb}\Z_{WZW}\Z_{gh},
\eeq{3.e}


where
\beqn
\Z_{ff} &=& \int \D\chi^{\dagger} \D\chi \exp( -\frac{1}{\pi}
\int(\chi^{\dagger}_2 \bar{\partial} \chi_1 +
\chi^{\dagger}_1 \partial \chi_2)d^2x), \nonumber \\
\Z_{fb} &=& \int \D\phi_a \D\phi_b
\exp(\frac{k}{\pi}\int \phi_a \Delta \phi_a d^2x+
\frac{k}{\pi}\int \phi_b \Delta \phi_b d^2x),
\nonumber
\\
\Z_{WZW} &=& \int \D \tilde U \exp\left((2k+2)\Gamma [\tilde U]\right).
\eeqn{3.e'}
$\Gamma [\tilde U]$ is the WZW action \cite{Wi} (\ref{WZW}) for the
gauge invariant combination $\tilde U = U^{-1} V$,
and $Z_{gh}$ corresponds to the Fadeev-Popov ghosts partition
function, whose explicit form will not be needed.

The central charge is now easily evaluated as the sum of four
independent
contributions coming from the different sectors, $c_{ff}=2k$,
$c_{fb}=2$, $c_{WZW}=2(k+1)(k^2-1)/(k+2)$ and $c_{gh}=-2(k^2+1)$,
thus giving
\beq
c=\frac{2(k-1)}{k+2} ,
\eeq{a2}
which coincides with the central charge of the $Z_k$ parafermion model.

After decoupling the gauge field $b_{\mu}^3$ the components of the
fermions defined in (\ref{gif}) can be written as
\beq
\begin{array}{ll}
\hat{\psi_1}^i(z)
=e^{\varphi_b(z)}{\tilde{\psi}}_1^i(z)
& \hat{\psi_1}^{i \dagger}(\bar z) =
{\tilde{\psi}}_1^{i \dagger}(\bar z) e^{\bar{\varphi}_b(\bar z)}
\\
\hat{\psi}_2^i(\bar z) =
e^{-\bar{\varphi}_b(\bar z) }{\tilde{\psi}}_2^i(\bar z)
& \hat{\psi_2}^{i \dagger}(z)
={\tilde{\psi}}_2^{i \dagger}(z) e^{\varphi_b(z)}  ,
\end{array}
\eeq{3.3}
where  $\tilde{\psi}$ stands for the fermions decoupled from the gauge
field $b_{\mu}$, and
\beq
\begin{array}{l}
\varphi_b (z)= \phi_b+i\int_x^{\infty} dz_{\mu} \epsilon_{\mu\nu}
\partial_{\nu}\phi_b \\
\bar{\varphi}_b (\bar z)= \phi_b-i\int_x^{\infty} dz_{\mu}
\epsilon_{\mu\nu}
\partial_{\nu}\phi_b
\end{array}
\eeq{3.4}
are the chiral (holomorphic and anti-holomorphic) components of the free
boson $\phi_b$.

We are now in position to evaluate the conformal dimensions and the
algebra obeyed by the fields defined in (\ref{sm}) and (\ref{smg}).

In fact, in the completely decoupled picture the fields defined in
(\ref{sm}) can be written as
\beqn
\sigma_1&\equiv & {\hat g}_{1,1}=
:\chi_2^{1\alpha} {\tilde U}^{-1\ \alpha\beta}
\chi_2^{\dagger 1\beta}: :e^{2\phi_a}::e^{2\phi_b}: +\ {\it h.c.}\
,
\nonumber \\
\mu_1&\equiv & {\hat g}^{\dagger}_{2,1}=
:\chi_1^{2\alpha} {\tilde U}^{\alpha\beta}
\chi_1^{\dagger 1\beta}:
:e^{-2\phi_a}::e^{\varphi_b-{\bar \varphi}_b}: +\ {\it h.c.}\ ,
\eeqn{smdesac}
and their conformal dimensions are then given by the sum of
the contributions
$h=1/2$, coming from free fermions, $h=-(k^2-1)/(2k(k+2))$,
coming from the WZW field $\tilde g$,
and twice $h=-1/4k$, coming from the vertex operators of the free
bosons $\phi_a$ and $\phi_b$.
This sum leads to the result quoted in (\ref{dim}).

For general values of $p$, eq.(\ref{smg}), the decoupled expression is
\beq
\begin{array}{l}
\sigma_p\equiv
:e^{2p\phi_a}::e^{2p\phi_b}:
:\chi_2^{1\alpha_1}...\chi_2^{1\alpha_p}::\chi_2^{\dagger 1\gamma_1}...
\chi_2^{\dagger 1\gamma_p}:
:{\tilde{U}}_{\cal A}^{-1\ \alpha_1\gamma_1...\alpha_p\gamma_p}: +\
{\it h.c.}\ ,
\nonumber \\
\mu_p\equiv
:e^{-2p\phi_a}::e^{p(\varphi_b-{\bar{\varphi}}_b)}:
:\chi_1^{2\alpha_1}...\chi_1^{2\alpha_p}::\chi_1^{\dagger 1\gamma_1}...
\chi_1^{\dagger 1\gamma_p}:
:{\tilde U}_{\cal A}^{\alpha_1\gamma_1...\alpha_p\gamma_p}: 
+\ {\it h.c.}\  ,
\end{array}
\eeq{smgdesac}
where
\beq
U_{\cal A}^{\alpha_1\gamma_1...\alpha_r\gamma_r}\equiv
[:U^{\alpha_1\gamma_1}\ldots U^{\alpha_r\gamma_r}:]_{\cal A}
\eeq{ga}
and the subscript ${\cal A}$ means antisymmetrization of the left and
right indices.

The conformal dimensions of $\sigma_p$ and $\mu_p$
are now given by the sum of
the contributions
$h=p/2$, coming from free fermions, $h=-p(k-p)(k+1)/(2k(k+2))$,
coming from the WZW field ${\tilde{U}}_{\cal A}$
and twice $h=-p/4k$, coming from the vertex operators. The result
coincides with eq.(\ref{dimen}).

The order/disorder algebras (\ref{od}) and (\ref{odg}) are easily
calculated using the decoupled expressions,
(\ref{smdesac}) and (\ref{smgdesac}), and making
use of the canonical commutation relations of the decoupled fields
\cite{CR1}.

In fact, this algebra has its origin in the odd
combinations of the
holomorphic and anti-holomorphic components of the free boson $\phi_b$
in eqs.(\ref{smdesac}) and (\ref{smgdesac}). Indeed, one can easily
check that
\beq
:e^{2p\phi_b(x_1)}::e^{p'(-\varphi_b+{\bar{\varphi}}_b)(x_2)}:
=e^{\frac{i2pp'\pi}{N}\Theta(x_1-x_2)}
:e^{p'(-\varphi_b+{\bar{\varphi}}_b)(x_2)}::e^{2p\phi_b(x_1)}:   ,
\eeq{origin}
being the other factors commuting.

Finally, the fields defined in (\ref{pfc}) are given
in terms of the decoupled fields by the expressions
\beqn
\psi_1^{PF}&\equiv & {\hat{\psi}}_1^2 {\hat{\psi}}_2^{1\dagger}=
:e^{-2\varphi_b}: \chi_1^2 \chi_2^{1\dagger} +\ {\it h.c.}\ ,
\nonumber \\
{\bar \psi}_1^{PF}&\equiv &
{\hat \psi}_2^2{\hat \psi}_1^{1\dagger}=
:e^{-2{\bar \varphi}_b}: \chi_2^2 \chi_1^{1\dagger} +\ {\it h.c.}
\eeqn{pfa}
Their conformal dimensions are  easily evaluated and give the result
quoted in eq.(\ref{dimpf}).
For the more general parafermion currents, eq.(\ref{pfg}), the evaluation of
the conformal dimensions proceeds similarly.

It is straightforward, from the decoupled expressions to derive the
whole set of conformal operator product expansions quoted in eq.(\ref{cope}).


\begin{thebibliography}{99}
\bibitem{FK} E.~Fradkin and L.P.~Kadanoff, Nucl.~Phys. {\bf B170} (1980)
1.

\bibitem{FZ} A.B.~Zamolodchikov and V.A.~Fateev, Sov.~Phys.~JETP
{\bf 62} (1985) 215.

\bibitem{F} V.A.~Fateev, Int.~J.~Mod.~Phys.~A {\bf6} (1991) 2109.

\bibitem{FZ2} V.A.~Fateev and Al.B.~Zamolodchikov, Phys.~Lett. {\bf B271}
(1991) 91.

\bibitem{Bakas} I.~Bakas, Int.~J.~Mod.Phys.~A {\bf 9} (1994) 3443.

\bibitem{Park} Q-H.~Park, Phys.~Lett. {\bf B328} (1994) 329 .

\bibitem{PS} Q-H.~Park and H.J.~Shin, Phys. ~Lett. {\bf B359} (1995) 125.

\bibitem{NS} S.G.~Naculich and H.J.~Schnitzer, Nucl.~Phys. {\bf B332}
(1990) 583.

\bibitem{BRS} K.~Bardakci, E.~Rabinovici and B.~S\"aring, Nucl.~Phys.
{\bf B299} (1988) 151.

\bibitem{nos} D.C.~Cabra, E.F.~Moreno and C.~von~Reichenbach,
Int.~J.~Mod.~Phys.~A {\bf 5} (1990) 2313.

\bibitem{GW} D.~Gepner and E.~Witten, Nucl.~Phys. {\bf B278} (1986) 493.

\bibitem{FGK} G.~Felder, K.~Gawdezki and A.~Kupiainen, Commun.~Math.~Phys.
{\bf 117} (1988) 83.

\bibitem{BCR} K.~Bardakci, M.~Crescimanno and E.~Rabinovici,
Nucl.~Phys. {\bf B344} (1990) 344; Nucl.~Phys. {\bf B349} (1991) 439.

\bibitem{CR1} D.C.~Cabra and K.D.~Rothe, Phys.~Rev.~D {\bf 51} (1995) R2509.

\bibitem{CR2} D.C.~Cabra and K.D.~Rothe, ``Fermionic Coset Realization of
Primaries in Critical Statistical Models'', hep-th 9503110, Ann.~Phys.
to appear.

\bibitem{FiPol} A.~Polyakov and P.~Wiegmann, Phys.~Lett. {\bf 131B}
(1983) 121; R.E.~Gamboa Sarav\'\i\, F.A.~Schaposnik and J.E.~Solomin,
Nucl.~Phys. {\bf B185} (1981) 238.

\bibitem{Wi} E.~Witten, Commun.~Math.~Phys. {\bf 92} (1984) 455.

\bibitem{So} O.~Soloviev, ``Interacting Wess-Zumino-Novikov-Witten models'',
QMW-PH-95-26, hep-th 9507004.

\end{thebibliography}
\end{document}